\documentclass[a4paper,11pt]{article}

\usepackage{jheppub}
\usepackage{orcidlink}
\usepackage{amsmath,bm}
\usepackage{amssymb,amsfonts}
\usepackage{lmodern}
\usepackage{parskip}
\usepackage{dsfont}

\begin{document}
\title{Path Integral Approach to Quantum Fisher Information}
\author[a]{Francis J. Headley\,\orcidlink{0009-0000-7585-4957},}
\emailAdd{francis.headley@uni-tuebingen.de}
\author[a]{Mahdi RouhbakhshNabati\,\orcidlink{0009-0006-3288-3372},}
\emailAdd{mahdi@rouhbakhsh.net}
\author[b]{Henry Harper-Gardner,}

\author[a]{Daniel Braun\,\orcidlink{0000-0001-8598-2039}}

\author[d]{Henning Schomerus\,\orcidlink{0000-0002-7959-0992},}
\author[a,c]{Emre Köse\,\orcidlink{0000-0001-9301-8709}}
\emailAdd{emrekose@ific.uv.es}

\affiliation[a]{Institut für Theoretische Physik, Eberhard-Karls-Universität Tübingen, 72076 Tübingen, Germany}
\affiliation[b]{Joint Quantum Centre (JQC) Durham-Newcastle,
School of Mathematics, Statistics and Physics, Newcastle University}
\affiliation[c]{Instituto de Física Corpuscular (IFIC), CSIC–Universitat de València, and Departament de Física Teòrica, Universitat de València, Parc Científic UV, c/ Catedrático José Beltrán, 2, E-46980 Paterna (València), Spain}
\affiliation[d]{Department of Physics, Lancaster University, Lancaster, LA1 4YB, United Kingdom}

\abstract{We present a real-time path-integral formulation of the quantum Fisher information for dynamical parameter estimation. For pure states undergoing unitary evolution, we show that the quantum Fisher information can be expressed as a connected symmetrized covariance of a time-integrated action deformation, equivalently as an integrated insertion of $\partial_\lambda S$ in the propagator. This reformulation avoids explicit state reconstruction by rewriting the quantum Fisher information in terms of real-time correlators that are natural targets for many-body methods. We further embed the construction into the Schwinger–Keldysh closed-time-path formalism, identifying the quantum Fisher information with the Keldysh component of an appropriate contour-ordered correlator generated by forward and backward propagating sources. Finally, using the Van Vleck–Gutzwiller approximation we re-derive the compact semiclassical  quantum Fisher information expression, clarifying how classical trajectory data control leading-order metrological sensitivity.}

\maketitle

\newpage

\section{Introduction}

The fundamental goal of quantum and classical metrology is to determine the ultimate precision with which an unknown parameter $\lambda$ can be inferred from measurement outcomes. In classical estimation theory the classical Fisher information 
\begin{equation}
F_{\mathrm{cl}}(\lambda)=\int_{\mathbb{R}}dx\,p_\lambda(x)\left(\dfrac{\partial\ln p_\lambda(x)}{\partial\lambda}\right)^2,
\label{cfi}
\end{equation}
quantifies the amount of information about parameter $\lambda$ encoded into the probability distribution $p_\lambda(x)$ of a random variable $x$. In quantum mechanics, $\lambda$ parametrises a density matrix $\rho_\lambda$, which describes the state of the system, where now $x$ denotes the random measurement outcome with probability $p_\lambda(x)=\mathrm{Tr}[M_x\rho_\lambda]$, given by the Born rule. Here $M_x$ is the positive operator-valued measure (POVM) element associated with outcome $x$, with $\int dx\,M_x=\bm{1}$.
Optimizing $F_{\mathrm{cl}}(\lambda)$ over all POVMs $M_x$ yields the Quantum Fisher Information (QFI) \cite{helstrom_quantum_1969,Holevo1982,braunstein_statistical_1994}:
\begin{equation}
    F_q:=\mathrm{Tr}(\rho_\lambda L_\lambda^2),
\end{equation}
where the symmetric logarithmic derivative (SLD) $L_\lambda$ is defined via 
\begin{equation}
    \dfrac{\partial\rho_\lambda}{\partial\lambda}=\dfrac{1}{2}(L_\lambda \rho_\lambda+ \rho_\lambda L_\lambda).
\end{equation}
The smallest uncertainty (standard deviation) of an unbiased estimator of $\lambda$ is then bounded from below by the quantum Cram\'er–Rao bound \cite{Cramer46,Rao1945},
\begin{equation}
\delta \lambda\geq  \dfrac{1}{\sqrt{Q F_{\mathrm{cl}}(\lambda)}}\geq  \dfrac{1}{\sqrt{Q F_q(\rho_\lambda)}},
\end{equation}
where $Q$ is the number of experiments performed \cite{Holevo1982,braunstein_statistical_1994}. The measurement that achieves the QFI is generally a projective measurement onto the eigen-basis of the SLD, however we note that the form of these optimal measurements is not always straightforward to obtain and is often not experimentally accessible \cite{braunstein_statistical_1994,Holevo1982}. The QFI is therefore typically used as a benchmark: one computes $F_q$ and compares it to the classical Fisher information attainable with a restricted measurement set \cite{braun_ultimate_2011,albarelli_evaluating_2019,paris_quantum_2009,toth_quantum_2014,headley_quantum_2025}. 

In many-body physics the QFI is closely tied to the fidelity susceptibility and is widely used as a diagnostic of quantum criticality \cite{PhysRevE.74.031123,Zanardi2007,PhysRevA.75.032109,gu_fidelity_2010,mukherjee_fidelity_2011,ilias_criticality-enhanced_2022}. For ground states $|\psi_0(\lambda)\rangle$, the overlap $|\langle\psi_0(\lambda)|\psi_0(\lambda+\delta\lambda)\rangle|$ quantifies how rapidly the state changes under an infinitesimal parameter shift; near a critical point this sensitivity typically becomes singular and exhibits characteristic system-size scaling \cite{gu_fidelity_2010,ilias_criticality-enhanced_2022,hauke_measuring_2016}. More generally, for a one-parameter family of pure states $|\psi(\lambda)\rangle$ one has the expansion
\begin{equation}
|\langle\psi(\lambda)|\psi(\lambda+\delta\lambda)\rangle|^2
=1-\frac{1}{4}F_\lambda\,(\delta\lambda)^2+O(\delta\lambda^3),
\end{equation}
which makes explicit that $F_\lambda/4$ is the Bures/Fubini-Study metric controlling infinitesimal state distinguishability \cite{braunstein_statistical_1994,giovannetti_advances_2011}.\footnote{For the remainder of this work, $F_\lambda$ will denote the single-parameter QFI. The multi-parameter QFI will be denoted as $F_{ij}$.} An essentially equivalent object also appears in holography, where the ground-state metric is often termed the quantum information metric (QIM) \cite{miyaji_gravity_2015,miyaji_butterflies_2016,trivella_holographic_2017,bak_information_2016,dimov_holographic_2021}, and Lagrangian techniques have been employed in the context of Euclidean path integrals \cite{alvarez-jimenez_quantum_2017,juarez_generalized_2023}. In particular, the QFI has been related to its geometric bulk quantities, such as the canonical energy \cite{miyaji_gravity_2015, lashkari_canonical_2016}.

In practice, the pure-state QFI is most often computed from state derivatives (or equivalently from the SLD). This is routine in few-body problems, but quickly becomes expensive in large Hilbert spaces and in interacting many-body
quantum-field theory (QFT) 
settings, where eigenstates are unavailable and real-time dynamics is typically accessible only approximately. In these regimes, derivative- or overlap-based evaluations of $F_\lambda$ are often the dominant computational bottleneck.  
Calculations of QFI for QFTs have been performed in the formalism of second quantization for both bosons \cite{Pinel12,safranek_ultimate_2015,safranek_estimation_2018,safranek_simple_2018,Ahmadi14.2} and fermions \cite{giraud_quantum_2024}. 

In this work we reformulate the pure-state QFI for unitary parameter encoding directly in a real-time path-integral language. Starting from the overlap definition of the QFI, we show that when the parameter dependence enters through the action, the QFI can be written as the connected symmetrized covariance of a time-integrated deformation operator. Equivalently, in the path integral the relevant quantity is an insertion of $\partial_\lambda S$ generated by differentiating the propagator with respect to the parameter. This bypasses explicit reconstruction of the evolved state or the SLD and instead reduces metrological sensitivity to correlation functions that are standard targets for many-body approaches (diagrammatics, tensor networks for correlators, semiclassical stationary phase, etc.).

A key outcome is that the resulting correlator has a transparent interpretation in the Schwinger–Keldysh closed-time-path (CTP) formalism: the dynamical QFI is generated by mixed functional derivatives of a CTP generating functional with independent sources on the forward and backward branches, and is identified with the Keldysh (symmetrized) component of a contour-ordered correlator. This situates the QFI within the real-time framework used for nonequilibrium many-body physics, and clarifies how the relevant insertions appear in Schwinger–Keldysh language, which is the natural starting point for treating environments via influence functionals. For continuum QFTs the deformation operators are composite, and the spacetime-integrated correlators entering the QFI generally require the usual renormalisation; the present formulation makes explicit which insertions must be renormalised. Finally, we present how the path-integral expression reduces in the semiclassical limit to a compact variance formula controlled by classical trajectory data within the Van Vleck–Gutzwiller approximation, generalizing the result of Rouhbakhshnabati \emph{et al.} \cite{rouhbakhshnabati_semiclassical_2025}.

\section{Path integral quantum Fisher information}

The QFI of a pure state $|\psi(t)\rangle$ is defined through the SLD, which is given as:
\begin{equation}
    L_\lambda = 2(|\partial_\lambda \psi(t)\rangle\langle  \psi (t)| + | \psi(t)\rangle\langle \partial_\lambda \psi (t)|),
 \end{equation}
where the derivative acts as $|\partial_\lambda \psi(t)\rangle=\frac{\partial}{\partial\lambda}|\psi(t)\rangle$. In the case that only a single-parameter $\lambda$ is of interest, the single-parameter pure state QFI is written as
\begin{equation}
    F_\lambda=4(\langle \partial_\lambda \psi (t)|\partial_\lambda \psi(t)\rangle-|\langle \psi(t)|\partial_\lambda \psi(t)\rangle|^2) \label{qfi}.
\end{equation}
This can be easily generalised through the SLD formalism to multiple parameters, such that the multi-parameter pure state QFI matrix is:
\begin{equation}
    F_{ij}=4\mathrm{Re}(\langle \partial_i \psi (t)|\partial_j \psi(t)\rangle-\langle \psi(t)|\partial_i \psi(t)\rangle(\langle \partial_j \psi(t)|\psi(t)\rangle)) \label{mpqfi}.
\end{equation}
where $\partial_i=\partial/\partial\lambda_i$ for a set of parameters $\{\lambda_i\}$. For the remainder of this text we only treat the single-parameter case. To arrive at a path-integral 
expression we must first write the state $|\psi(t)\rangle$ in the Schrödinger representation and insist that the parameter dependence is encoded entirely in the unitary evolution:
\begin{equation}
    |\psi(t)\rangle=U_\lambda(t,0)\,|\psi(0)\rangle,
\end{equation}
where we also assume that the initial state is parameter independent. Here $U_\lambda(t,0)$ denotes the unitary propagator, which in general is generated by a time dependent Hamiltonian. For time-independent $\widehat{H}_\lambda$, this reduces to $U_{\lambda}(t,0)=e^{-i\widehat{H}_{\lambda}t/\hbar}$. When the Hamiltonian is explicitly time-dependent the evolution is generated by the time-ordered propagator
\begin{equation}
    U_\lambda(t,0)=T \exp{\left(-\frac{i}{\hbar}\int_0^t dt' \widehat{H}_\lambda(t')\right)}.
    \label{time_evo}
\end{equation}
Taking the second term of \eqref{qfi}, we insert resolutions of the identity in the position basis, $\bm{1}=\int dq |q\rangle\langle q|$, to express the overlap in terms of propagator kernels:
\begin{align}
    \langle \psi(t)|\partial_\lambda \psi(t)\rangle&=\int dq_2  \langle \psi(t)|q_2\rangle\langle q_2|\partial_\lambda \psi(t)\rangle\nonumber\\
    &=\int dq_2  \langle \psi(0)|U^\dagger_\lambda(t,0)|q_2\rangle\langle q_2|\partial_\lambda U_\lambda(t,0)|\psi(0)\rangle\nonumber\\
     &=\int dq_1 dq_2 dq_3  \langle \psi(0)|q_3\rangle\langle q_3|U^\dagger_\lambda(t,0)|q_2\rangle\left(\dfrac{\partial}{\partial\lambda}\langle q_2| U_\lambda(t,0)|q_1\rangle\right)\langle q_1|\psi(0)\rangle.
     \label{qfi2}
\end{align}
Although we will not derive the path integral here, we will note that the propagator terms may be expressed as integrals over all possible paths between two points\footnote{This is generally the case for Hamiltonians in which the momentum terms are at most quadratic.}, where \cite{Schulman1981,peskin}:
\begin{equation}
    \langle q_2| U_\lambda(t,0)|q_1\rangle= \int\displaylimits_{\tilde{x}(0)=q_1}^{\tilde{x}(t)=q_2} \mathcal{D}\tilde{x}\exp\left(\frac{i}{\hbar}S_\lambda [\tilde{x}]\right). \label{pathint1}
\end{equation}
In the path integral representation, the required time ordering is implemented automatically by the time-sliced construction, so the kernel representation Eq.~\eqref{pathint1} remains valid without modification (with $\int_0^t dt' L_\lambda(x,\dot x,t')$ when the Lagrangian is explicitly time-dependent).
\begin{equation}
    \langle q_2|T\exp\left[-\dfrac{i}{\hbar}\int\displaylimits_0^t dsH_\lambda(s)\right]|q_1\rangle=\int\displaylimits_{\tilde{x}(0)=q_1}^{\tilde{x}(t)=q_2} \mathcal{D}\tilde{x}\exp\left(\frac{i}{\hbar}S_\lambda [\tilde{x}]\right). \label{time_ordered}
\end{equation}
We emphasise that path integrals are implicitly time ordered objects \cite{Schulman1981,peskin}. With this in mind, the following results also hold for time-dependent Hamiltonians or Lagrangians.
Substituting \eqref{pathint1} into \eqref{qfi2}, we find a path integral expression for the second term of \eqref{qfi}:
\begin{equation}
    \langle \psi(t)|\partial_\lambda \psi(t)\rangle =\dfrac{i}{\hbar}\int dq_1\,dq_2\,dq_3\ \psi^*(0,q_3)\psi(0,q_1)\left(\ \int\displaylimits_{q_3}^{q_2} \mathcal{D}\tilde{x}\,e^{iS_\lambda [\tilde{x}]/\hbar}\right)^{\hspace{-1mm}*}\left(\ \int\displaylimits_{q_1}^{q_2} \mathcal{D}\tilde{y}\,e^{iS_\lambda [\tilde{y}]/\hbar}\dfrac{\partial S_\lambda}{\partial \lambda}\right), \label{qfit1}
\end{equation}
where we assume that the path integral measures $\mathcal{D}\tilde{x}$ and $\mathcal{D}\tilde{y}$, and the endpoints, are parameter independent. The same procedure may be applied to $\langle \partial_\lambda \psi (t)|\partial_\lambda \psi(t)\rangle$ to recover the following expression:
\begin{equation}
     \dfrac{1}{\hbar^2}\int dq_1\,dq_2\,dq_3\ \psi^*(0,q_3)\psi(0,q_1)\left(\ \int\displaylimits_{q_3}^{q_2} \mathcal{D}\tilde{x}e^{iS_\lambda [\tilde{x}]/\hbar}\dfrac{\partial S_\lambda}{\partial \lambda}\right)^{\hspace{-1mm}*}\left(\ \int\displaylimits_{q_1}^{q_2} \mathcal{D}\tilde{y}e^{iS_\lambda [\tilde{y}]/\hbar}\dfrac{\partial S_\lambda}{\partial \lambda}\right). \label{qfit2}
\end{equation}
Substituting \eqref{qfit1} and \eqref{qfit2} into \eqref{qfi} we may express the QFI of a pure state in the path integral formalism:
\begin{align}
    F_{\lambda}=\dfrac{4}{\hbar^2}&\left[\int dq_1\,dq_2\,dq_3\ \psi^*(0,q_3)\psi(0,q_1)\left(\ \int\displaylimits_{q_3}^{q_2} \mathcal{D}\tilde{x}e^{iS_\lambda [\tilde{x}]/\hbar}\dfrac{\partial S_\lambda}{\partial \lambda}\right)^{\hspace{-1mm}*}\left(\ \int\displaylimits_{q_1}^{q_2} \mathcal{D}\tilde{y}e^{iS_\lambda [\tilde{y}]/\hbar}\dfrac{\partial S_\lambda}{\partial \lambda}\right)\right.\nonumber\\
    &\quad \left. -\left|\int dq_1\,dq_2\,dq_3\ \psi^*(0,q_3)\psi(0,q_1)\left(\ \int\displaylimits_{q_3}^{q_2} \mathcal{D}\tilde{x}e^{iS_\lambda [\tilde{x}]/\hbar}\right)^{\hspace{-1mm}*}\left(\ \int\displaylimits_{q_1}^{q_2} \mathcal{D}\tilde{y}e^{iS_\lambda [\tilde{y}]/\hbar}\dfrac{\partial S_\lambda}{\partial \lambda}\right)\right|^2\right]. \label{pqfi}
\end{align}
The above result has only been derived for the single-parameter case. Although the multi-parameter QFIM can be derived similarly, the expressions are lengthy and can be found in the Appendix \ref{App: multi-para}.

The representation \eqref{pqfi} has been expressed in configuration space, but this structure extends naturally to continuum quantum fields and many-body systems. To see this, let $|\phi\rangle$ denote an eigenstate of the field operator $\hat\phi(\mathbf r)$
at fixed time, with eigenvalue configuration $\phi(\mathbf r)$. For a pure initial state $|\psi(0)\rangle$, define the wave-functional
\begin{equation}
\Psi_0[\phi] := \langle \phi | \psi(0)\rangle.
\end{equation}
The real-time propagator kernel in field-configuration space is given:
\begin{equation}
K_\lambda[\phi_f,\phi_i;t] := \langle \phi_f|U_\lambda(t,0)|\phi_i\rangle = \int\displaylimits_{\phi(0)=\phi_i}^{\phi(t)=\phi_f}\mathcal D\phi\;
e^{\frac{i}{\hbar}S_\lambda[\phi]}.
\end{equation}
The field-theoretic analogue of Eq.~\eqref{pqfi} is obtained using nearly identical steps, and the result is equivalent to replacing the ordinary integrals with functional integrals and the endpoint coordinates with field configurations. Thus, we have
\begin{align}
F_\lambda(t)=\frac{4}{\hbar^2}&\Bigg[
\int \hspace{-1mm}  D\phi_1\,  D\phi_2\,  D\phi_3
\Psi_0^*[\phi_3]\Psi_0[\phi_1]\!
\left(\int\displaylimits_{\,\,\phi(0)=\phi_3}^{\phi(t)=\phi_2}\hspace{-4mm}\mathcal D\phi\;
e^{\frac{i}{\hbar}S_\lambda[\phi]}\,
\dfrac{\partial S_\lambda}{\partial \lambda}\right)^{\hspace{-1mm}*}
\hspace{-1mm}\left(\int\displaylimits_{\,\,\phi'(0)=\phi_1}^{\phi'(t)=\phi_2}\hspace{-4mm}\mathcal D\phi'\;
e^{\frac{i}{\hbar}S_\lambda[\phi']}\,
\dfrac{\partial S_\lambda}{\partial \lambda}\right)
\nonumber\\
&\quad -
\left|
\int \hspace{-1mm}  D\phi_1\,  D\phi_2\,  D\phi_3
\Psi_0^*[\phi_3]\Psi_0[\phi_1]\!
\left(\int\displaylimits_{\,\,\phi(0)=\phi_3}^{\phi(t)=\phi_2}\hspace{-4mm}\mathcal D\phi\;
e^{\frac{i}{\hbar}S_\lambda[\phi]}\,
\right)^{\hspace{-1mm}*}\,
\hspace{-1mm}\int\displaylimits_{\phi'(0)=\phi_1}^{\phi'(t)=\phi_2}\hspace{-4mm}\mathcal D\phi'\;
e^{\frac{i}{\hbar}S_\lambda[\phi']}\,
\dfrac{\partial S_\lambda}{\partial \lambda}
\right|^2
\Bigg],\label{eq:F_field_exact}
\end{align}
where $\int \mathcal{D}\phi$ denotes a path integral over field histories with fixed endpoints, while $\int D\phi_i$ denotes a functional integral over field configurations. This is our main result. 

We now specialize to the case where the parameter enters through a local deformation of the Lagrangian density, such that the $\lambda$-dependence of the action is simply
\begin{equation}
S_\lambda[\phi]=\int_0^{t}\!dt'\int d^3r\,\mathcal L_\lambda[\phi],
\qquad
\partial_\lambda S_\lambda[\phi]=\int_0^{t}\!dt'\int d^3r\,\partial_\lambda \mathcal L_\lambda[\phi].
\end{equation}
Here we also assume a standard real-time configuration-space path integral with $\lambda$-independent measure and endpoint constraints, so that $\partial_\lambda$ acts only on the phase $e^{\frac{i}{\hbar}S_\lambda[x]}$ (equivalently on $\partial_\lambda\mathcal L_\lambda$), and does not generate additional Jacobian or boundary contributions (see Appendix~\ref{app:insertion}). A convenient operator quantity to associate with $\partial_\lambda S_\lambda$ is the standard Hermitian generator of the unitary family
$|\psi_\lambda(t)\rangle = U_\lambda(t,0)|\psi(0)\rangle$,
\begin{equation}
\widehat{G}_\lambda(t)\;:=\; i\,U_\lambda^\dagger(t,0)\,\partial_\lambda U_\lambda(t,0),
\qquad \widehat{G}_\lambda^\dagger=\widehat{G}_\lambda .
\label{eq:Gdef}
\end{equation}
For a pure state the quantum Fisher information is the variance of this Hermitian generator,
\begin{equation}
F_\lambda(t)=4\Big(\langle \widehat{G}_\lambda(t)^2\rangle-\langle \widehat{G}_\lambda(t)\rangle^2\Big),
\qquad
\langle\cdot\rangle\equiv \langle\psi(0)|\cdot|\psi(0)\rangle .
\label{eq:QFIvarG}
\end{equation}
This is the standard ``variance of a Hermitian generator'' form; compared to the path-integral insertion $(i/\hbar)\,\partial_\lambda S_\lambda$, we have simply stripped the overall factor $i/\hbar$ and packaged the result into the Hermitian $\widehat{G}_\lambda(t)$. To connect $\widehat{G}_\lambda(t)$ to a time-integrated deformation operator, we use the exact Duhamel identity
\begin{equation}
\widehat{G}_\lambda(t)=\frac{1}{\hbar}\int_0^{t}\!dt'\;\Big[U_\lambda^\dagger(t',0)\,(\partial_\lambda \widehat{H}_\lambda(t'))\,U_\lambda(t',0)\Big]
\equiv \frac{1}{\hbar}\int_0^{t}\!dt'\,(\partial_\lambda \widehat{H}_\lambda(t'))_H .
\label{eq:Gduhamel}
\end{equation}
For \emph{coupling} deformations (i.e.\ when $\lambda$ enters only as a coupling multiplying a term in the Lagrangian density and does not modify the kinetic structure), one has at fixed canonical variables
\begin{equation}
\partial_\lambda \widehat H_\lambda(t)
= -\int d^3 r\,\partial_\lambda \widehat{\mathcal L}_\lambda(t,\mathbf r).
\label{eq:dH_dlambda_fixed}
\end{equation}
The local Lagrangian density operator is defined by promoting the classical fields to canonical operators:
\begin{equation}
\hat{\mathcal{L}}_\lambda(\mathbf x) = \mathcal{L}_\lambda(\hat{\phi}(\mathbf x), \nabla\hat{\phi}(\mathbf x), \hat{\pi}(\mathbf x)).
\end{equation}
Here $\hat{\pi}(x)$ is the canonical momentum density operator conjugate to the field satisfying the standard equal-time commutation relation
$[\hat{\phi}(\mathbf x), \hat{\pi}(\mathbf y)] = i\hbar\delta^{(3)}(\mathbf{x}-\mathbf{y})$. It is natural to define the Hermitian local deformation operator in the Schr\"odinger picture,
 \begin{equation}
 \hat o \;\equiv\; \int d^3x\,\partial_\lambda\widehat{\mathcal L}_\lambda(\mathbf x),
 \label{eq:odef_S}
 \end{equation}
as well as its Heisenberg-picture counterpart
\begin{equation}
\hat o_{H}(t) \;\equiv\; U_\lambda^\dagger(t,0)\,\hat o\,U_\lambda(t,0)
\;=\; \int d^3x\,\partial_\lambda\widehat{\mathcal L}_\lambda(t, \mathbf x)\Big|_{H},
\label{eq:odef_H}
\end{equation}
where the subscript $H$ denotes Heisenberg evolution with $\widehat{H}_\lambda$.  For the remainder of this work, all operators carrying an explicit time argument are understood to be in the Heisenberg picture unless stated otherwise. The deformation generator \eqref{eq:Gduhamel} then reads
\begin{equation}
\widehat{G}_\lambda(t)= -\frac{1}{\hbar}\int_0^{t}\!dt'\,\hat o_{H}(t').
\label{eq:G_from_oH}
\end{equation}
Substituting Eq.~\eqref{eq:G_from_oH} into \eqref{eq:QFIvarG} and expanding the variance gives a compact real-time expression in terms of a connected symmetrized
correlator,
\begin{equation}
F_\lambda(t)
=\frac{4}{\hbar^2}
\int_0^{t}\!dt'\int_0^{t}\!dt''\;\frac{1}{2}
\Big\langle \big\{\delta \hat o_{H}(t'),\delta \hat o_{H}(t'')\big\}\Big\rangle,
\qquad
\delta \hat o \equiv \hat o-\langle \hat o\rangle ,
\label{eq:QFIsymm}
\end{equation}
which matches Eq.~(5) of Ilias \textit{et al.}~\cite{ilias_criticality-enhanced_2022} (see also
\cite{gammelmark_fisher_2014,hauke_measuring_2016}). This correlator is precisely the Keldysh (symmetrized) component of the Schwinger-Keldysh two-point function of $\hat o$.
We make the observation that \eqref{eq:QFIsymm} may be written as a Keldysh path integral. But to make the closed-time-path structure explicit, we must first introduce the standard Schwinger-Keldysh (CTP) generating functional with independent sources $J_\pm$ on the forward/backward branches \cite{bentov_schwinger-keldysh_2021}:
\begin{align} Z_\lambda[J_+,J_-]
&= \int \mathcal{D}\phi_f\,\mathcal{D}\phi_i\,\mathcal{D}\phi_i'\;
\rho_0(\phi_i,\phi_i')
\int\displaylimits_{\phi_+(0)=\phi_i}^{\phi_+(t)=\phi_f} \!\mathcal{D}\phi_+
\int\displaylimits_{\phi_-(0)=\phi_i'}^{\phi_-(t)=\phi_f} \!\mathcal{D}\phi_-
\nonumber\\
&\quad \times
\exp\!\left\{
\frac{i}{\hbar}\Big(S_\lambda[\phi_+]-S_\lambda[\phi_-]\Big)
+ \frac{i}{\hbar}\int_0^{t}\!dt'\,
\Big(J_+(t')\,o[\phi_+(t')] - J_-(t')\,o[\phi_-(t')]\Big)
\right\},
\label{eq:CTP_Z}
\end{align}
where the fields $ \phi_+(t)$ and $ \phi_-(t)$, integrated over the common final configuration, satisfy the boundary condition $\phi_\pm(t)=\phi_f$ and are the fields on the forward/backward branches of the contour $C$.
For a pure state $\rho_0\!\,\big( \phi_+(0), \phi_-(0)\big) = \Psi_0[\phi_+] \Psi_0^*[\phi_-]$. Functional differentiation of the CTP generating functional \eqref{eq:CTP_Z} generates branch-resolved correlators. To obtain compact expressions, we defined the source-dependent expectation value
\begin{align}
\langle \mathcal A\rangle_{J}
:=\frac{1}{Z_\lambda[J_+,J_-]}&\int \mathcal{D}\phi_f\,\mathcal{D}\phi_i\,\mathcal{D}\phi_i'\;
\rho_0(\phi_i,\phi_i')
\int\displaylimits_{\phi_+(0)=\phi_i}^{\phi_+(t)=\phi_f} \!\mathcal{D}\phi_+
\int\displaylimits_{\phi_-(0)=\phi_i'}^{\phi_-(t)=\phi_f} \!\mathcal{D}\phi_-A[ \phi_+, \phi_-]\nonumber\\
&\quad\times
e^{
\frac{i}{\hbar}\Big(S_\lambda[\phi_+]-S_\lambda[\phi_-]\Big)
+ \frac{i}{\hbar}\int_0^{t}\!dt'\,
\Big(J_+(t')\,o[\phi_+(t')] - J_-(t')\,o[\phi_-(t')]\Big)}
.
\label{eq:expJ_def}
\end{align}
A straightforward differentiation shows (see Appendix~\ref{app:CTP_derivatives}) that the mixed second derivative of $\ln Z_\lambda$ yields the connected Wightman correlator,
\begin{equation}
\frac{\delta^2 \ln Z_\lambda}{\delta J_-(t')\,\delta J_+(t'')}\Bigg|_{J_\pm=0}
=
\frac{1}{\hbar^2}\Big(\langle \hat o_{H}(t')\hat o_{H}(t'')\rangle-\langle \hat o_{H}(t')\rangle\langle \hat o_{H}(t'')\rangle\Big),
\label{eq:mixed_lnZ_identity}
\end{equation}
with an analogous identity for $\delta^2\ln Z/(\delta J_+(t')\delta J_-(t''))$ obtained by exchanging the operator order. Summing the two mixed derivatives therefore reproduces the symmetrized connected correlator appearing in \eqref{eq:QFIsymm},
\begin{equation}
\frac12\Big\langle\big\{\delta \hat o_{H}(t'),\delta \hat o_{H}(t'')\big\}\Big\rangle
=\frac{\hbar^2}{2}\left[
\frac{\delta^2 \ln Z_\lambda}{\delta J_-(t')\,\delta J_+(t'')}
+
\frac{\delta^2 \ln Z_\lambda}{\delta J_+(t')\,\delta J_-(t'')}
\right]_{J_\pm=0},
\label{eq:symm_from_lnZ}
\end{equation}
Therefore, the dynamical QFI can be written directly as a functional-derivative:

\begin{equation}
F_\lambda(t) = 4\int_0^{t}\!dt'\int_0^{t}\!dt''\;
\left[
\frac{\delta^2 \ln Z_\lambda[J_+,J_-]}{\delta J_-(t')\,\delta J_+(t'')}\right]_{J_\pm=0},
\label{eq:QFI_from_lnZ}
\end{equation}
where we have used the fact that the integration domain is symmetric under $t'\leftrightarrow t''$. We emphasize that Eq. \eqref{eq:QFI_from_lnZ} has been derived for pure states under unitary parameter encoding. However for open dynamics, influence functionals provide a convenient representation of reduced density matrices and measurement statistics on the system \cite{feynman_vernon_1963, kading_new_2023}. When the environment is incorporated through an influence functional so that $Z_\lambda[J_+,J_-]$ generates the system correlators of the underlying system–environment unitary dilation, Eq. \eqref{eq:QFI_from_lnZ} can be used to evaluate the corresponding global  QFI, which is the same ‘global QFI’ appearing in the continuous-measurement formulations of Gammelmark/Ilias \cite{gammelmark_fisher_2014,ilias_criticality-enhanced_2022}. We note however that a genuinely reduced-state QFI would require a separate mixed-state (Bures/Uhlmann) construction.

\section{Semiclassical reduction from the field theory} \label{sec:SC}

In the work of RouhbakhshNabati \emph{et al.} \cite{rouhbakhshnabati_semiclassical_2025} the authors presented a semiclassical approximation of the QFI in configuration space. We refer the reader to Appendix \ref{sec:semiclassical_from_pqfi_1d} for more information on the original configuration space derivation. Here we generalise this result and present a field-theoretic semiclassical extension of the QFI, starting from the expression \eqref{eq:F_field_exact} and utilising the field theory versions of the Van Vleck-Gutzwiller (VVG) propagator and Wigner function. We first assume that the real-time path integral yields a semiclassical stationary phase expansion about its classical solutions of the Euler–Lagrange field equations with fixed boundary conditions
\begin{equation}
\phi_\gamma(\mathbf r,0)=\phi_i(\mathbf r),\qquad
\phi_\gamma(\mathbf r,t)=\phi_f(\mathbf r),
\end{equation}
where $\gamma$ denotes the different paths of the field evolution. Formally we may write our field theory VVG stationary-phase approximation 
\cite{CotlerWei2023QuantumScarsQFT}: 
\begin{equation}
K_\lambda[\phi_f,\phi_i;t]
\approx
\sum_{\gamma:\,\phi_i\to\phi_f}
A_\gamma[\phi_f,\phi_i;t]\,
\exp\!\left[
\frac{i}{\hbar}S_\gamma[\phi_f,\phi_i;t;\lambda]
-\frac{i\pi}{2}\nu_\gamma
\right].
\label{eq:K_sc_field_cont}
\end{equation}
Here $S_\gamma[\phi_f,\phi_i;t;\lambda]$ is the classical action evaluated on the solution $\gamma$, $\nu_\gamma$ is the Maslov index, and $A_\gamma[\phi_f,\phi_i;t]$ is the semiclassical amplitude arising from quadratic fluctuations about $\gamma$. $A_\gamma$ is chosen real and non-negative; all semiclassical phases from Gaussian fluctuations (including caustics) are absorbed into the Maslov index $\nu_\gamma$ in Eq.~\eqref{eq:K_sc_field_cont}. For a lattice or finite-mode regularisation with $N$ (configuration-space) degrees of freedom, the amplitude takes the explicit form
\begin{equation}
A_\gamma[\phi_f,\phi_i;t]
=(2\pi \hbar)^{-N/2}
\left|\det\nolimits \left(
\frac{\delta^2 S_\gamma[\phi_f,\phi_i;t;\lambda]}
{\delta\phi_f(\mathbf r')\,\delta \phi_i(\mathbf r)}\right)\right|^{1/2},
\label{eq:A_detM_precise}
\end{equation}
where the functional determinant is understood as a regulated functional determinant of the quadratic fluctuation operator. For a lattice or finite-mode cutoff it reduces to an ordinary finite-dimensional determinant; for regulators such as dimensional regularisation or smearing it remains a functional determinant. The precise numerical prefactor in \eqref{eq:A_detM_precise} is regularisation-scheme dependent; in a lattice regularisation with $N$ sites it reduces to the standard finite-dimensional Van Vleck normalisation $(2\pi\hbar)^{-N/2}$, while other schemes may redistribute overall constants between the amplitude and the functional measure. Upon removing the regulator, the determinant and prefactor must be defined within the chosen regularisation/renormalisation scheme, but the variance formula remains well-defined provided the Wigner functional is properly normalised. In the finite-dimensional case (for lattices or a finite-mode cutoff), one recovers Appendix~\ref{sec:semiclassical_from_pqfi_1d} with $N=r$.

The endpoint functional derivatives of the classical action define the canonical momentum fields at the initial and final times,
\begin{equation}
\pi_{f,\gamma}(\mathbf r)
=
\frac{\delta S_\gamma[\phi_f,\phi_i;t;\lambda]}{\delta \phi_f(\mathbf r)},
\qquad
\pi_{i,\gamma}(\mathbf r)
=
-\frac{\delta S_\gamma[\phi_f,\phi_i;t;\lambda]}{\delta \phi_i(\mathbf r)}.
\label{eq:endpoint_action_field_cont}
\end{equation}
Differentiating the semiclassical propagator \eqref{eq:K_sc_field_cont} with respect to $\lambda$ gives
\begin{equation}
\partial_\lambda K_\lambda
\approx
\sum_\gamma
A_\gamma
e^{\frac{i}{\hbar}S_\gamma-\frac{i\pi}{2}\nu_\gamma}
\left(
\frac{i}{\hbar}\,\partial_\lambda S_\gamma
+\partial_\lambda \ln A_\gamma
\right),
\label{eq:dK_sc_field_cont_full}
\end{equation}
where the $\lambda$-dependence of $S_\gamma$ and $A_\gamma$ is implicit. Away from caustics (or conjugate points), and under the standard semiclassical hierarchy 
\begin{equation}
\partial_\lambda S_\gamma = O(\hbar^0),\qquad
\partial_\lambda \ln A_\gamma = O(\hbar^0),
\end{equation}
the phase derivative term dominates by an explicit factor of $1/\hbar$. We therefore retain only the leading term
\begin{equation}
\partial_\lambda K_\lambda[\phi_f,\phi_i;t]
\approx
\frac{i}{\hbar}
\sum_\gamma
\big(\partial_\lambda S_\gamma\big)\,
A_\gamma[\phi_f,\phi_i;t]\,
\exp\!\left[
\frac{i}{\hbar}S_\gamma[\phi_f,\phi_i;t;\lambda]
-\frac{i\pi}{2}\nu_\gamma
\right].
\label{eq:dK_sc_field_cont_leading}
\end{equation}
Substituting Eqs.~\eqref{eq:K_sc_field_cont} and \eqref{eq:dK_sc_field_cont_leading} into Eq. \eqref{eq:F_field_exact}, and keeping the diagonal (stationary-phase) contribution as in the finite-dimensional derivation, one obtains 
\begin{align}
\left(\partial_\lambda K\right)^*\left(\partial_\lambda K\right)
&\;\approx\;
\frac{1}{\hbar^2}\sum_{\gamma}
\left(\partial_\lambda S_\gamma\right)^2
\left|A_\gamma\right|^2\,\exp{(\frac{i}{\hbar}\Delta S_\gamma)}, \label{eq:schemat1}
\\[4pt]
(K)^*\left(\partial_\lambda K\right)
&\;\approx\;
\frac{i}{\hbar}\sum_{\gamma}
\left(\partial_\lambda S_\gamma\right)
\left|A_\gamma\right|^2\,\exp{(\frac{i}{\hbar}\Delta S_\gamma)}, \label{eq:schemat2}
\end{align}
where
\begin{equation}
\Delta S_\gamma
=
S_\gamma[\phi_2,\phi_1;t;\lambda]
-
S_\gamma[\phi_2,\phi_3;t;\lambda].
\label{eq:DeltaS_field_cont}
\end{equation}
We introduce centre and relative initial field configurations,
\begin{equation}
\Phi(\mathbf r)
:=
\frac{\phi_1(\mathbf r)+\phi_3(\mathbf r)}{2},
\qquad
\xi(\mathbf r)
:=
\phi_1(\mathbf r)-\phi_3(\mathbf r).
\label{eq:Phi_xi_def_cont}
\end{equation}
for which the functional measure transforms formally as
\begin{equation}
D\phi_1\,D\phi_3
=
D\Phi\,D\xi,
\label{eq:measure_center_relative_cont}
\end{equation}
up to a field-independent Jacobian absorbed into the overall normalization. We then expand the action difference \eqref{eq:DeltaS_field_cont} to first order in the relative field $\xi$ at fixed $\phi_2$. This is a functional Taylor expansion with respect to the initial endpoint field $\phi_i(\mathbf r)$, so the linear term carries an integration over the spatial coordinate:
\begin{align}
\Delta S_\gamma
&=
S_\gamma\!\left[\phi_2,\Phi+\frac{\xi}{2};t;\lambda\right]
-
S_\gamma\!\left[\phi_2,\Phi-\frac{\xi}{2};t;\lambda\right]
\nonumber\\
&\approx
\int d^3r\;
\xi(\mathbf r)\,
\left.
\frac{\delta S_\gamma[\phi_2,\phi_i;t;\lambda]}{\delta \phi_i(\mathbf r)}
\right|_{\phi_i=\Phi}\nonumber\\
&
=
-\int d^3r\;\xi(\mathbf r)\,\Pi_\gamma(\mathbf r;\phi_2,\Phi,t),
\label{eq:DeltaS_expand_cont}
\end{align}
where we have defined
\begin{equation}
    \Pi_\gamma(\mathbf r)\equiv-\left.\frac{\delta S_\gamma[\phi_2,\phi_i;t;\lambda]}{\delta \phi_i(\mathbf r)}\right|_{\phi_i=\Phi}
\end{equation}
as the initial canonical momentum field of the specific classical branch $\gamma$ connecting $\Phi$ to $\phi_2$ in time $t$. Accordingly, the oscillatory phase factor in Eqs.~\eqref{eq:schemat1} and \eqref{eq:schemat2} becomes
\begin{equation}
\exp\!\left(\frac{i}{\hbar}\Delta S_\gamma\right)
\approx
\exp\!\left[
-\frac{i}{\hbar}\int d^3r\;\Pi_\gamma(\mathbf r)\,\xi(\mathbf r)
\right].
\label{eq:phase_pi_xi_cont}
\end{equation}
The phase in Eq.~\eqref{eq:phase_pi_xi_cont} has precisely the Fourier structure that appears in the definition of the field-theoretic Wigner functional. As such we introduce
\begin{equation}
\Psi_0^*\!\left[\Phi-\frac{\xi}{2}\right]
\Psi_0\!\left[\Phi+\frac{\xi}{2}\right]
\end{equation}
to produce the field-theoretic Wigner functional of the initial state:
\begin{equation}
W_0[\Phi,\Pi]
:=
\mathcal N_W
\int D\xi\;
\exp\!\left[
-\frac{i}{\hbar}\int d^3r\;\Pi(\mathbf r)\,\xi(\mathbf r)
\right]
\Psi_0^*\!\left[\Phi-\frac{\xi}{2}\right]
\Psi_0\!\left[\Phi+\frac{\xi}{2}\right],
\label{eq:Wigner_functional_def_cont}
\end{equation}
where $\Pi$ is, at this stage, a dummy phase-space argument (not yet identified with the branch-dependent momentum field $\Pi^\gamma$). The normalization factor $\mathcal N_W$ is chosen such that
\begin{equation}
\int D\Phi\,D\Pi\;W_0[\Phi,\Pi]=1.
\label{eq:Wigner_functional_norm_cont}
\end{equation}
To convert the remaining integration over the endpoint field configuration $\phi_2$ into an integration over initial momenta, one proceeds branch-by-branch. For a fixed classical branch $\gamma$, the semiclassical amplitude and endpoint measure combine through the functional analogue of the Van Vleck Jacobian, 
\begin{equation}
|A_\gamma[\phi_2,\Phi;t]|^2\,D\phi_2
\;=\;
(2\pi \hbar)^{-N}
\,\left|\det\!\left(\frac{\delta \Pi_\gamma}{\delta \phi_2}\right)\right|
D\phi_2
=
\frac{D\Pi_\gamma}{(2\pi \hbar)^{N}}.
\label{eq:measure_conversion_functional_cont}
\end{equation}

After this branchwise change of variables, the initial momentum field becomes the phase-space integration variable; at that stage we suppress the branch label and write it simply as $\Pi$. This leads to the semiclassical phase-space functional average
\begin{equation}
\langle G\rangle_{\rm sc}
:=
\int D\Phi\,D\Pi\;
W_0[\Phi,\Pi]\,
G[\Phi,\Pi],
\label{eq:sc_average_functional_cont}
\end{equation}
where $G[\Phi,\Pi]$ is evaluated on the classical field solution $\phi_{\rm cl}$ generated by the initial data
\begin{equation}
\phi_{\rm cl}(\mathbf r,0)=\Phi(\mathbf r),
\qquad
\pi_{\rm cl}(\mathbf r,0)=\Pi(\mathbf r).
\end{equation}
In this way, the finite-dimensional semiclassical construction of Ref.~\cite{rouhbakhshnabati_semiclassical_2025} is lifted to a continuum field-theoretic phase-space formulation: the ordinary Wigner function is replaced by a Wigner functional, and the finite-dimensional Van Vleck determinant is replaced by a regulated functional Jacobian. With this notation, the leading-order semiclassical field-theoretic QFI takes the same variance form as in the finite-dimensional case,
\begin{equation}
F_\lambda(t)
\;\approx\;
\frac{4}{\hbar^2}
\left(
\left\langle\left(\partial_\lambda S_{\rm cl}\right)^2\right\rangle_{\rm sc}
-
\left\langle\partial_\lambda S_{\rm cl}\right\rangle_{\rm sc}^2
\right)
=
\frac{4}{\hbar^2}\,
\mathrm{Var}_{\rm sc}\!\left(\partial_\lambda S_{\rm cl}\right),
\label{eq:F_sc_variance_field}
\end{equation}
where $\partial_\lambda S_{\rm cl}$ denotes the parametric derivative of the classical action evaluated on the classical field solution generated by the initial data $(\Phi,\Pi)$,
\begin{equation}
\partial_\lambda S_{\rm cl}[\Phi,\Pi]
=
\partial_\lambda S_\lambda[\phi_{\rm cl}]
=
\int_0^t dt'\int d^3r\;
\partial_\lambda \mathcal L_\lambda\!\big(\phi_{\rm cl}(\mathbf r,t'),\partial\phi_{\rm cl}(\mathbf r,t');\mathbf r,t'\big).
\label{eq:dS_generator_field}
\end{equation}
Here $\phi_{\rm cl}$ is the solution of the classical Euler-Lagrange equations with initial conditions
\begin{equation}
\phi_{\rm cl}(\mathbf r,0)=\Phi(\mathbf r),
\qquad
\pi_{\rm cl}(\mathbf r,0)=\Pi(\mathbf r).
\label{eq:field_initial_data_sc}
\end{equation}
Equation~\eqref{eq:F_sc_variance_field} is the continuum field-theoretic analogue of the semiclassical result of Ref.~\cite{rouhbakhshnabati_semiclassical_2025}: the QFI is determined, at leading semiclassical order, by the fluctuations of the accumulated action deformation $\partial_\lambda S_{\rm cl}$ over the classical ensemble weighted by the initial-state Wigner functional.

\section{Discussion}

The path-integral representation derived here provides access to the QFI in settings where direct operator-based constructions (e.g.\ via explicit SLD eigenstructures) become impractical. For interacting many-body systems and quantum field theories where exact eigenstates are unavailable, Eq.~\eqref{pqfi} rewrites the QFI in terms of standard real-time correlation functions generated by an insertion $\partial_\lambda S$. In practice, this opens the door to controlled approximations: in perturbation theory $\partial_\lambda S$ acts as an additional vertex, and semiclassical stationary-phase methods apply when $\hbar\to0 $. For continuum QFTs and coupling deformations, Eq.~\eqref{eq:F_field_exact} expresses the QFI in terms of spacetime-integrated connected correlators of the deformation operator. In an interacting QFT these are correlators of composite operators and therefore generally require renormalisation, including possible contact terms and associated scheme dependence. While a full analysis is beyond the scope of the present note, the formulation makes transparent which operator insertions control the QFI and thereby provides a clean starting point for applying standard composite-operator renormalisation methods. By formulating the quantum Fisher information in a field theoretic framework, this work may be extended to further study model discrimination of different QFTs, and in particular enable studies which allow us to better understand experimental sensing of fifth forces, dark matter and other conjectured high-energy phenomena.

The real-time structure also interfaces naturally with nonequilibrium techniques. Extending beyond pure, unitary dynamics (for example to mixed states or open systems) can be approached via purification/Bures-metric formulations or via channel-QFI constructions; in a Schwinger-Keldysh description this generically leads to contour-ordered objects involving influence functionals and response kernels. A detailed development of these extensions (including the precise choice of mixed-state QFI notion and the role of non-unitary dynamics) is left for future work, but the present representation clarifies the contour/operator structures one would need to generalise. The semiclassical limit (Sec.~\ref{sec:SC}) provides a transparent physical interpretation: the QFI reduces to the variance of accumulated action derivatives over classical trajectory ensembles, consistent with recent applications to chaotic dynamics \cite{rouhbakhshnabati_semiclassical_2025}. Furthermore, the equivalence to fidelity susceptibility (for coupling deformations) suggests applications to quantum critical phenomena, where universal scaling can be analysed via renormalisation-group flows near fixed points \cite{Sachdev2011,gu_fidelity_2010}. Overall, the path-integral formulation provides a practical bridge between metrological distinguishability measures and the standard toolkit of nonequilibrium and many-body theory, particularly in regimes where state-vector methods are cumbersome.

\section{Conclusion}

We have derived a real-time path-integral representation of the quantum Fisher information for pure states under unitary parameter encoding. The formulation expresses the QFI as a connected quadratic functional of the action insertion $\partial_\lambda S$, making explicit how parameter distinguishability 
accumulates through time evolution. For parameters entering as coupling constants, this reduces to spacetime-integrated connected two-point functions of the deformation operator, establishing equivalence with fidelity-susceptibility formulas used in many-body physics \cite{ilias_criticality-enhanced_2022,hauke_measuring_2016}. 
This clarifies the structural aspects of quantum estimation and emphasizes the role of temporal correlations.

Because the QFI is written in terms of standard contour-ordered correlators and action insertions, the framework extends directly to field-theoretic methods, such as diagrammatic expansions, semiclassical approximations, and composite-operator renormalisation. This positions the representation as a useful bridge between quantum metrology and many-body theory, enabling computation and benchmarking of ultimate sensitivity to couplings, masses, and interaction strengths in systems where explicit operator/SLD constructions are impractical.

As an application of the formalism, we replicated the result that under a stationary-phase approximation, the QFI reduces to the variance of classical action derivatives over trajectory ensembles, connecting quantum and classical notions of distinguishability \cite{rouhbakhshnabati_semiclassical_2025}, and demonstrated that this connection also extends to the field-theoretic context.
Further natural applications include time-dependent settings, such as quenches and driving, where the time-contour formulation remains valid and can capture transient sensitivity enhancements. 
The path-integral approach also opens a route toward incorporating noise and decoherence through influence functionals, leading to a unified treatment of quantum Fisher information in open quantum systems.  
Practical settings that can benefit from this field-theoretic formulation range from criticality-enhanced sensing to semiclassical gravity and high-energy contexts, where action-based formulations are often the most natural starting point.

\section*{Acknowledgements}
We thank Malik Jirasek, Viktoria Noel, Gabrielle Perfetto, Shozab Qasim, and Alexander Jahn for fruitful discussions. Furthermore we thank Christian Käding for his feedback on the draft. FJH and DB acknowledge the EU EIC Pathfinder project QuCoM (101046973). HHG was funded by the UKRI and EPSRC (2902876). MR was funded by the Deutsche Forschungsgemeinschaft (DFG, German Research Foundation), Project No. 465199066. EK acknowledges the project PID2023-152724NA-I00, with funding from MCIU/AEI / 10.13039/501100011033 and FSE+, by the project CNS2024-154818 with funding from MICIU/AEI /10.13039/501100011033.

\bibliographystyle{apsrev4-2}
\bibliography{refs.bib}

\appendix
\newpage
\section{Path-integral insertion as an operator} \label{app:insertion}

We begin from the insertion amplitude
\begin{equation}
  C_\mathcal{O}(q_f,t)
  \;:=\;
  \int dq_i\,\psi(0,q_i)
  \int_{x(0)=q_i}^{x(t)=q_f}\!\!\mathcal{D}x\;
  e^{\tfrac{i}{\hbar}S[x]}\,\mathcal{O}[x],
  \label{eq:Cf-def}
\end{equation}
where $\mathcal{O}[x]$ is an arbitrary functional built from time-local insertions along
$[0,t]$. Throughout we assume the standard configuration-space path-integral
representation with $\lambda$-independent measure and $\lambda$-independent
endpoint constraints; otherwise additional Jacobian/boundary contributions can
appear.

We now regard the path integral with insertion $\mathcal{O}[x]$ as defining the kernel of
an operator $\hat {\mathcal{O}}(t)$ by
\begin{equation}
  \langle q_f|\hat{\mathcal{O}}(t)|q_i\rangle
  \;:=\;
  \int_{x(0)=q_i}^{x(t)=q_f}\!\!\mathcal{D}x\;
  e^{\tfrac{i}{\hbar}S[x]}\,\mathcal{O}[x],
  \label{eq:Af-kernel}
\end{equation}
in direct analogy with the usual identification of the bare propagator kernel
\begin{equation}
  \langle q_f|U(t,0)|q_i\rangle
  \;=\;
  K(q_f,t;q_i,0)
  \;=\;
  \int_{x(0)=q_i}^{x(t)=q_f}\!\!\mathcal{D}x\;
  e^{\tfrac{i}{\hbar}S[x]}.
  \label{eq:K-kernel}
\end{equation}
With this definition, Eq.~\eqref{eq:Cf-def} can be written compactly as
\begin{equation}
  C_\mathcal{O}(q_f,t)
  \;=\;
  \int dq_i\,\psi(0,q_i)\,\langle q_f|\hat {\mathcal{O}}(t)|q_i\rangle
  \;=\;
  \langle q_f|\hat {\mathcal{O}}(t)|\psi(0)\rangle,
  \label{eq:Cf-operator}
\end{equation}
where we used $|\psi(0)\rangle=\int dq_i\,|q_i\rangle\,\psi(0,q_i)$.

\subsection*{Specialization to $\mathcal{O}[x]=\partial_\lambda S_\lambda$}

In the case relevant for the QFI we take $\mathcal{O}[x]=\partial_\lambda S_\lambda[x]$,
where $\lambda$ is the parameter to be estimated and all $\lambda$-dependence
enters through $S_\lambda[x]$. The $\lambda$-dependent propagator is
\begin{equation}
  K_\lambda(q_f,t;q_i,0)
  \;=\;
  \int_{q_i}^{q_f}\mathcal{D}x\;
  e^{\tfrac{i}{\hbar}S_\lambda[x]}
  \;=\;
  \langle q_f|U_\lambda(t,0)|q_i\rangle.
\end{equation}
Differentiating with respect to $\lambda$ yields
\begin{equation}
  \partial_\lambda K_\lambda(q_f,t;q_i,0)
  \;=\;
  \int_{q_i}^{q_f}\mathcal{D}x\;
  e^{\tfrac{i}{\hbar}S_\lambda[x]}\,
  \frac{i}{\hbar}\,\partial_\lambda S_\lambda[x]
  \;=\;
  \frac{i}{\hbar}\,
  \langle q_f|\hat {\mathcal{O}}(t)|q_i\rangle,
\end{equation}
and hence
\begin{equation}
  \langle q_f|\hat{\mathcal{O}}(t)|q_i\rangle
  \;=\;
  \frac{\hbar}{i}\,\partial_\lambda
  \langle q_f|U_\lambda(t,0)|q_i\rangle.
\end{equation}
Therefore, at the operator level,
\begin{equation}
  \hat{\mathcal{O}}(t)
  \;=\;
  \frac{\hbar}{i}\,\partial_\lambda U_\lambda(t,0).
  \label{eq:A-dU}
\end{equation}
Using Eq. \eqref{eq:Gdef}, $\hat O(t)=-\hbar\, U_\lambda(t,0)\widehat{G}_\lambda(t)$, so $\widehat{G}_\lambda(t)=-\frac{1}{\hbar}\, U_\lambda^\dagger(t,0)\hat O(t)$. Using Eq.~\eqref{eq:Cf-operator}, we immediately obtain
\begin{equation}
  C_\mathcal{O}(q_f,t)
  \;=\;
  \langle q_f|\hat{\mathcal{O}}(t)|\psi(0)\rangle
  \;=\;
  \frac{\hbar}{i}\,
  \langle q_f|\partial_\lambda U_\lambda(t,0)|\psi(0)\rangle
  \;=\;
  \frac{\hbar}{i}\,\partial_\lambda\psi_\lambda(t,q_f),
  \label{eq:C-dpsi}
\end{equation}
where $\psi_\lambda(t,q_f)=\langle q_f|\psi_\lambda(t)\rangle$ and
$|\psi_\lambda(t)\rangle=U_\lambda(t,0)|\psi(0)\rangle$.

\section{Keldysh Generating Functional}
\paragraph{CTP generating functional and mixed second derivatives.}\label{app:CTP_derivatives}

We start from the closed-time-path (CTP) generating functional
\begin{align}
Z_\lambda[J_+,J_-]
&= \int \mathcal{D}\phi_f\,\mathcal{D}\phi_i\,\mathcal{D}\phi_i'\;
\rho_0(\phi_i,\phi_i')
\int\displaylimits_{\phi_+(0)=\phi_i}^{\phi_+(t)=\phi_f} \!\mathcal{D}\phi_+
\int\displaylimits_{\phi_-(0)=\phi_i'}^{\phi_-(t)=\phi_f} \!\mathcal{D}\phi_-
\nonumber\\
&\quad \times
\exp\!\left\{
\frac{i}{\hbar}\Big(S_\lambda[\phi_+]-S_\lambda[\phi_-]\Big)
+ \frac{i}{\hbar}\int_0^{t}\!dt'\,
\Big(J_+(t')\,o[\phi_+(t')] - J_-(t')\,o[\phi_-(t')]\Big)
\right\}.
\label{eq:CTP_Z_recall}
\end{align}

Define the expectation values with sources by
\begin{align}
\langle \mathcal A\rangle_{J}
:=\frac{1}{Z_\lambda[J_+,J_-]}&\int \mathcal{D}\phi_f\,\mathcal{D}\phi_i\,\mathcal{D}\phi_i'\;
\rho_0(\phi_i,\phi_i')
\int\displaylimits_{\phi_+(0)=\phi_i}^{\phi_+(t)=\phi_f} \!\mathcal{D}\phi_+
\int\displaylimits_{\phi_-(0)=\phi_i'}^{\phi_-(t)=\phi_f} \!\mathcal{D}\phi_-A[ \phi_+, \phi_-]\nonumber\\
&\quad\times
e^{
\frac{i}{\hbar}\Big(S_\lambda[\phi_+]-S_\lambda[\phi_-]\Big)
+ \frac{i}{\hbar}\int_0^{t}\!dt'\,
\Big(J_+(t')\,o[\phi_+(t')] - J_-(t')\,o[\phi_-(t')]\Big)}
.
\label{eq:exp_def}
\end{align}
 In particular $Z_\lambda[0,0]=\text{Tr}\rho_0$ (and equals $1$ if $\rho_0$ is normalized). Differentiating \eqref{eq:CTP_Z_recall} with respect to $J_+(t)$ gives
\begin{align}
\frac{\delta Z_\lambda}{\delta J_+(t)}
&=\int \mathcal{D}\phi_f\,\mathcal{D}\phi_i\,\mathcal{D}\phi_i'\;
\rho_0(\phi_i,\phi_i')
\int\displaylimits_{\phi_+(0)=\phi_i}^{\phi_+(t)=\phi_f} \!\mathcal{D}\phi_+
\int\displaylimits_{\phi_-(0)=\phi_i'}^{\phi_-(t)=\phi_f} \!\mathcal{D}\phi_-\\
&
\qquad\times\frac{\delta}{\delta J_+(t)}
e^{\frac{i}{\hbar}\Big(S_\lambda[\phi_+]-S_\lambda[\phi_-]\Big)
+ \frac{i}{\hbar}\int_0^{t}\!dt'\,
\Big(J_+(t')\,o[\phi_+(t')] - J_-(t')\,o[\phi_-(t')]\Big)}\nonumber\\
&=\int \mathcal{D}\phi_f\,\mathcal{D}\phi_i\,\mathcal{D}\phi_i'\;
\rho_0(\phi_i,\phi_i')
\int\displaylimits_{\phi_+(0)=\phi_i}^{\phi_+(t)=\phi_f} \!\mathcal{D}\phi_+
\int\displaylimits_{\phi_-(0)=\phi_i'}^{\phi_-(t)=\phi_f} \!\mathcal{D}\phi_-
\left(\frac{i}{\hbar}\,o[ \phi_+(t)]\right)\exp\left\{\cdots\right\}\;\nonumber\\
&=\frac{i}{\hbar}\,Z_\lambda[J_+,J_-]\;\langle o_+(t)\rangle_{J},
\label{eq:dZ_dJplus}
\end{align}
where $\exp\left\{\dots\right\}$ denotes the exponential with its weight and  $o_+(t)\equiv o[ \phi_+(t)]$. Dividing by $Z_\lambda$ gives
\begin{equation}
\frac{\delta \ln Z_\lambda}{\delta J_+(t)}=\frac{1}{Z_\lambda}\frac{\delta Z_\lambda}{\delta J_+(t)}
=\frac{i}{\hbar}\,\langle o_+(t)\rangle_{J}.
\label{eq:dlnZ_dJplus}
\end{equation}
Similarly, differentiating with respect to $J_-(t)$ yields an extra minus sign because the source appears as $-J_-o_-$:
\begin{align}
\frac{\delta Z_\lambda}{\delta J_-(t)}=-\frac{i}{\hbar}\,Z_\lambda[J_+,J_-]\;\langle o_-(t)\rangle_{J},
\end{align}
hence
\begin{equation}
\frac{\delta \ln Z_\lambda}{\delta J_-(t)}=-\frac{i}{\hbar}\,\langle o_-(t)\rangle_{J}.
\label{eq:dlnZ_dJminus}
\end{equation}
Differentiating \eqref{eq:dlnZ_dJplus} with respect to $J_-(t)$ gives
\begin{equation}
\frac{\delta^2 \ln Z_\lambda}{\delta J_-(t)\,\delta J_+(t')}
=\frac{i}{\hbar}\,\frac{\delta}{\delta J_-(t)}\langle o_+(t')\rangle_{J}.
\label{eq:start_mixed}
\end{equation}
Straightforward derivation yields
\begin{align}
\frac{\delta}{\delta J_-(t)}\langle o_+(t')\rangle_J
&=-\frac{i}{\hbar}\Big(\langle o_-(t)o_+(t')\rangle_J-\langle o_-(t)\rangle_J\langle o_+(t')\rangle_J\Big).
\label{eq:derivative_expectation}
\end{align}
Substituting \eqref{eq:derivative_expectation} into \eqref{eq:start_mixed} gives
\begin{equation}
\frac{\delta^2 \ln Z_\lambda}{\delta J_-(t)\,\delta J_+(t')}
=
\frac{1}{\hbar^2}\Big(\langle o_-(t)o_+(t')\rangle_J-\langle o_-(t)\rangle_J\langle o_+(t')\rangle_J\Big).
\label{eq:mixed_lnZ_generalJ}
\end{equation}
Finally set $J_\pm=0$:
\begin{equation}
\frac{\delta^2 \ln Z_\lambda}{\delta J_-(t)\,\delta J_+(t')}\Bigg|_{J_\pm=0}
=
\frac{1}{\hbar^2}\Big(\langle o_-(t)o_+(t')\rangle-\langle o_-(t)\rangle\langle o_+(t')\rangle\Big)
\equiv \frac{1}{\hbar^2}\,\langle o_-(t)o_+(t')\rangle_c.
\label{eq:mixed_lnZ_at0}
\end{equation}
At $J_\pm=0$, the CTP path integral reproduces contour-ordered operator expectation values.
In particular, because any point on the $(-)$ branch is later on the contour than any point on the $(+)$ branch,
\begin{equation}
\langle o_-(t)o_+(t')\rangle
=\langle T_C\,\hat o_C(t_-)\hat o_C(t'_+)\rangle
=\langle \hat o_H(t)\,\hat o_H(t')\rangle,
\label{eq:minusplus_is_wightman}
\end{equation}
so the mixed derivative generates the connected Wightman correlator:
\begin{equation}
\frac{\delta^2 \ln Z_\lambda}{\delta J_-(t)\,\delta J_+(t')}\Bigg|_{J_\pm=0}
=\frac{1}{\hbar^2}\Big(\langle \hat o_H(t)\hat o_H(t')\rangle-\langle \hat o_H(t)\rangle\langle \hat o_H(t')\rangle\Big).
\label{eq:mixed_lnZ_final}
\end{equation}

\section{Semiclassical reduction and the Van Vleck-Gutzwiller approximation}
\label{sec:semiclassical_from_pqfi_1d}

We now recap how the semiclassical QFI follows directly from the path-integral expression \eqref{pqfi} for a system with $r$ canonical degrees of freedom  as seen in \cite{rouhbakhshnabati_semiclassical_2025}. The key input is the stationary-phase approximation (Van Vleck-Gutzwiller) to the real-time propagator \cite{Schulman1981,Gutzwiller1990}. For a particle moving in this space the semiclassical propagator from $q_i$ to $q_f$ in time $t$ is
\begin{equation}
K^{\rm SC}_\lambda(q_f,q_i;t)\equiv \langle q_f|U_\lambda(t,0)|q_i\rangle
\;\approx\sum_{\gamma:\,q_i\to q_f}
A_\gamma(q_f,q_i;t)\,
\exp\!\left[\frac{i}{\hbar}S_\gamma(q_f,q_i;t;\lambda)
-\frac{i\pi}{2}\nu_\gamma\right],
\label{eq:VVG_1d}
\end{equation}
where $\gamma$ labels classical trajectories satisfying the boundary conditions \cite{Schulman1981,Gutzwiller1990}, $S_\gamma$ is the
(classical) action evaluated on the trajectory $\gamma$,
\begin{equation}
S_\gamma(q_f,q_i;t;\lambda)=\int_{0}^{t} dt'\,L_\lambda(q(t'),\dot{q}(t')),
\end{equation}
and $\nu_\gamma$ is the Maslov index \cite{Schulman1981,Maslov1972}. Then, the Van Vleck amplitude reads
\begin{equation}
A_\gamma(q_f,q_i;t)
=
(2\pi\hbar)^{-r/2}\,
\left|D_\gamma(q_f,q_i;t)\right|^{1/2},
\quad
D_\gamma(q_f,q_i;t)
:=
\det\left(-\frac{\partial^2 S_\gamma(q_f,q_i;t;\lambda)}{\partial q_f\,\partial q_i}\right),
\label{eq:VVdet_1d}
\end{equation}
with $D_\gamma$ the \emph{Van Vleck determinant} of a $r\times r$ matrix \cite{Schulman1981,Gutzwiller1990}. Using the standard endpoint relations (fixed $t$),
\begin{equation}
    p_f=\frac{\partial S_\gamma(q_f,q_i;t;\lambda)}{\partial q_f}, \qquad p_i=-\frac{\partial S_\gamma(q_f,q_i;t;\lambda)}{\partial q_i},
    \label{eq:pi_pf_def}
\end{equation}
one obtains the equivalent endpoint-stability forms
\begin{equation}
D_\gamma(q_f,q_i;t)=\det\left(-\frac{\partial p_f}{\partial q_i}\right)
=\det\left(\frac{\partial p_i}{\partial q_f}\right),
\label{eq:VVdet_1d_alt}
\end{equation}
where $p_{i,f}$ are the classical momenta at the initial/final endpoints of $\gamma$. Thus $|D_\gamma|^{1/2}$ measures the sensitivity of the classical boundary-value problem to its endpoint data. We assume a standard semiclassical hierarchy in which $\partial_\lambda S_\gamma=O(\hbar^0)$ and $ \partial_\lambda\ln A_\gamma=O(\hbar^0)$, so that in $\partial_\lambda K^{\rm SC}_\lambda$ the contribution from $(i/\hbar)\,(\partial_\lambda S_\gamma)K^{\rm SC}_\lambda$ dominates over $(\partial_\lambda\ln A_\gamma)K^{\rm SC}_\lambda$, away from caustics. When integrating over the intermediate coordinate in products of propagators, we employ a diagonal (stationary-phase) approximation in which the dominant contributions arise from pairings of trajectories with matching classical actions; off-diagonal terms are rapidly oscillatory and are suppressed provided the initial wavepacket is sufficiently localized and times are not so long that exponentially many distinct classical paths contribute. Near conjugate points/caustics, or in strongly multi-path regimes, amplitude-derivative and off-diagonal corrections can become important and the following expressions should be interpreted as leading-order semiclassical results. In \eqref{pqfi} the insertions $\partial_\lambda S$ arise through derivatives of propagators with respect to $\lambda$. Differentiating \eqref{eq:VVG_1d} gives
\begin{equation}
\partial_\lambda K^{\rm SC}_\lambda(q_f,q_i;t)
\;\approx\;
\sum_{\gamma}
A_\gamma e^{\frac{i}{\hbar}S_\gamma-\frac{i\pi}{2}\nu_\gamma}
\left(
\frac{i}{\hbar}\,\partial_\lambda S_\gamma
+\partial_\lambda \ln A_\gamma
\right).
\label{eq:dK_full_1d}
\end{equation}
The leading semiclassical contribution is the term proportional to $1/\hbar$. To leading order we therefore
neglect $\partial_\lambda\ln A_\gamma$ and keep only
\begin{equation}
\partial_\lambda K^{\rm SC}_\lambda(q_f,q_i;t)
\;\approx\;
\frac{i}{\hbar}\sum_{\gamma}
\left(\partial_\lambda S_\gamma\right)
A_\gamma(q_f,q_i;t)
\exp\!\left[\frac{i}{\hbar}S_\gamma(q_f,q_i;t;\lambda)
-\frac{i\pi}{2}\nu_\gamma\right].
\label{eq:dK_leading_1d}
\end{equation}
Here $\partial_\lambda S_\gamma$ is the parametric derivative of the classical action along the trajectory. Substituting Eqs. \eqref{eq:VVG_1d} and \eqref{eq:dK_leading_1d} into the QFI expression \eqref{pqfi} yields double sums over pairs of trajectories $\gamma,\gamma'$ from the complex-conjugated and non-conjugated propagators, with oscillatory phases $\exp[\tfrac{i}{\hbar}(S_\gamma-S_{\gamma'})]$. After integrating over the intermediate coordinate $q_2$, the leading semiclassical contribution comes from \emph{diagonal} pairings $\gamma=\gamma'$ (or more generally from stationary-phase pairings), for which the phase cancels.
Keeping only this diagonal contribution yields, schematically,
\begin{align}
\left(\partial_\lambda K^{\rm SC}\right)^*\left(\partial_\lambda K^{\rm SC}\right)
&\;\Rightarrow\;
\frac{1}{\hbar^2}\sum_{\gamma}
\left(\partial_\lambda S_\gamma\right)^2
\left|A_\gamma\right|^2\,\exp{(\frac{i}{\hbar}\Delta S_\gamma)}, \label{schemat1}
\\[4pt]
(K^{\rm SC})^*\left(\partial_\lambda K^{\rm SC}\right)
&\;\Rightarrow\;
\frac{i}{\hbar}\sum_{\gamma}
\left(\partial_\lambda S_\gamma\right)
\left|A_\gamma\right|^2\,\exp{(\frac{i}{\hbar}\Delta S_\gamma)}. \label{schemat2}
\end{align}
where the action difference in \eqref{pqfi} is
\begin{equation}
    \Delta S_\gamma=S_\gamma(q_2,q_1;t;\lambda)- S_\gamma(q_2,q_3;t;\lambda). \label{delta_S}
\end{equation}
We introduce center coordinate for the initial endpoints, to reduce the rest calculations to a phase-space average,
\begin{equation}
    q:=\frac{q_1+q_3}{2},\quad \xi:=q_1-q_3,\quad d^rq_1\,d^rq_3=d^rq\,d^r\xi.
\end{equation}
For the stationary-phase contribution the initial endpoints are close, $q_1\simeq q_3$, implying $\xi\ll 1$. Then, for a fixed $q_2$ we can expand Eq.~\eqref{delta_S} to leading order,
\begin{align}
    S_\gamma(q_2,q+\xi/2;t;\lambda)-S_\gamma(q_2,q-\xi/2;t;\lambda)&\approx\xi\cdot \nabla_{q'}S_\gamma(q_2,q';t;\lambda)|_{q'=q}\nonumber\\
    &=-\xi\cdot p_i(q_2,q;t;\lambda),
\end{align}
where $p_i(q_2,q;t;\lambda)$ is the initial momentum of the classical trajectory. From this we see that we can now write the integral in terms of the Wigner function of the initial state,
\begin{align}
    W_0(q,p):=\frac{1}{(2\pi\hbar)^r}\int d^{r}\xi\;e^{-\frac{i}{\hbar}p\cdot\xi}\,\psi_0^*(q-\xi/2)\,\psi_0(q+\xi/2),\quad \int d^{r}q\,d^{r}p\,W_0(q,p)=1.
\end{align}
The squared VVG propagator amplitude provides the Jacobian that transforms the integration over final coordinates into an integration over initial momenta,
\begin{align}
|A_\gamma(q_f,q_i;t)|^2 d^rq_f=
\frac{1}{(2\pi\hbar)^r}\,|D_\gamma(q_f,q_i;t)| d^rq_f=
\frac{1}{(2\pi\hbar)^r}\left|\det\left(\frac{\partial p_i}{\partial q_f}\right)\right| d^rq_f=\frac{d^rp_i}{(2\pi\hbar)^r}.
\end{align}
We can now define the compact semiclassical average of any trajectory functional $G$ as a Wigner-weighted phase-space average,
\begin{equation}
    \langle G \rangle_{\rm sc}:=\int d^rq\,d^rp\,G(q,p)\,W_0(q,p),
\end{equation}
where $G(q,p)$ is evaluated along the classical trajectory generated from the initial phase-space point $(q,p)$.
\footnote{Concretely, $\langle\cdot\rangle_{\rm sc}$ corresponds to integrating over $(q_i,q_f)$ with weights $\psi^*(0,q_3)\psi(0,q_1)$ and the semiclassical transition weights $|A_\gamma(q_f,q_i;t)|^2$ arising from the diagonal approximation. For minimum-uncertainty Gaussian packets, this can be equivalently rewritten as a Gaussian average over initial phase-space points centred at the packet centroid.}\\
With this notation, the semiclassical QFI to leading order becomes
\begin{equation}
F_\lambda(t)
\;\approx\;
\frac{4}{\hbar^2}
\left(
\left\langle\left(\partial_\lambda S_{\rm cl}\right)^2\right\rangle_{\rm sc}
-
\left\langle\partial_\lambda S_{\rm cl}\right\rangle_{\rm sc}^2
\right)
=
\frac{4}{\hbar^2}\,
\mathrm{Var}_{\rm sc}\!\left(\partial_\lambda S_{\rm cl}\right),
\label{eq:F_sc_variance_1d}
\end{equation}
where $\partial_\lambda S_{\rm cl}$ is the classical action derivative evaluated along the corresponding classical trajectory $\gamma$. This is exactly the result of \cite{rouhbakhshnabati_semiclassical_2025}. Equation \eqref{eq:F_sc_variance_1d} is precisely the semiclassical statement that the QFI is governed by the fluctuations (over the classical ensemble associated with the initial wavepacket) of the dynamical generator $\partial_\lambda S$ accumulated along classical motion.

\section{Multi-parameter generalisation} \label{App: multi-para}

In the main text we derived the single-parameter QFI in a path-integral and Schwinger-Keldysh language.  Here we present the extension to the multi-parameter QFIM.

\subsection*{Pure-state QFIM and its path-integral form}

Consider a set of parameters $\boldsymbol\lambda = (\lambda_1,\dots,\lambda_M)$ encoded in the unitary evolution $U_{\boldsymbol\lambda}(t,0)$.  The multi-parameter pure-state QFIM is \cite{braunstein_statistical_1994}
\begin{equation}
F_{ij}
= 4\,\mathrm{Re}\!\Big(
\langle \partial_i\psi(t)|\partial_j\psi(t)\rangle
-\langle\psi(t)|\partial_i\psi(t)\rangle\,
\langle\partial_j\psi(t)|\psi(t)\rangle
\Big),
\label{eq:QFIM_def_app}
\end{equation}
where $\partial_i \equiv \partial/\partial\lambda_i$.  Following exactly the same steps as in Section~2 -- inserting resolutions of the identity, expressing the propagator as a path integral, and noting that $\partial_i$ acts only on the phase $e^{(i/\hbar)S_{\boldsymbol\lambda}}$ -- we obtain the multi-parameter path-integral QFI:
\begin{align}
F_{ij}=\frac{4}{\hbar^2}\,\mathrm{Re}&\Bigg[
\int dq_1\,dq_2\,dq_3\;\psi^*(0,q_3)\,\psi(0,q_1)\,
\left(\int\displaylimits_{q_3}^{q_2}\mathcal D\tilde x\;
e^{iS_{\boldsymbol\lambda}[\tilde x]/\hbar}\,
\frac{\partial S_{\boldsymbol\lambda}}{\partial\lambda_i}\right)^{\!*}
\left(\int\displaylimits_{q_1}^{q_2}\mathcal D\tilde y\;
e^{iS_{\boldsymbol\lambda}[\tilde y]/\hbar}\,
\frac{\partial S_{\boldsymbol\lambda}}{\partial\lambda_j}\right)
\nonumber\\[4pt]
&\;-\left(
\int dq_1\,dq_2\,dq_3\;\psi^*(0,q_3)\,\psi(0,q_1)\,
\left(\int\displaylimits_{q_3}^{q_2}\mathcal D\tilde x\;
e^{iS_{\boldsymbol\lambda}[\tilde x]/\hbar}\right)^{\!*}
\left(\int\displaylimits_{q_1}^{q_2}\mathcal D\tilde y\;
e^{iS_{\boldsymbol\lambda}[\tilde y]/\hbar}\,
\frac{\partial S_{\boldsymbol\lambda}}{\partial\lambda_i}\right)
\right)
\nonumber\\[4pt]
&\;\times\left(
\int dq_1\,dq_2\,dq_3\;\psi^*(0,q_3)\,\psi(0,q_1)\,
\left(\int\displaylimits_{q_3}^{q_2}\mathcal D\tilde x\;
e^{iS_{\boldsymbol\lambda}[\tilde x]/\hbar}\,
\frac{\partial S_{\boldsymbol\lambda}}{\partial\lambda_j}\right)^{\!*}
\left(\int\displaylimits_{q_1}^{q_2}\mathcal D\tilde y\;
e^{iS_{\boldsymbol\lambda}[\tilde y]/\hbar}\right)
\right)
\Bigg].
\label{eq:QFIM_PI}
\end{align}
Note the appearance of $\mathrm{Re}(\cdots)$: whereas the single-parameter QFI~\eqref{pqfi} is manifestly real (the imaginary part vanishes identically for $i=j$), the off-diagonal elements of the QFIM require the explicit real part.

\subsection*{Deformation generators and variance form}

For each parameter $\lambda_i$ we define the Hermitian deformation generator by the multi-parameter Duhamel identity,
\begin{equation}
\widehat G_i(t)
\;:=\;
i\,U_{\boldsymbol\lambda}^\dagger(t,0)\,\partial_i U_{\boldsymbol\lambda}(t,0)
\;=\;
\frac{1}{\hbar}\int_0^t\!dt'\;
\big(\partial_i\widehat H_{\boldsymbol\lambda}(t')\big)_{\!H},
\label{eq:Gi_def}
\end{equation}
which generalises Eq.~\eqref{eq:Gdef}.  The QFIM then takes the covariance form
\begin{equation}
F_{ij}(t)
= 2\Big(
\big\langle\widehat G_i(t)\,\widehat G_j(t)\big\rangle
+
\big\langle\widehat G_j(t)\,\widehat G_i(t)\big\rangle
\Big)
-4\,\big\langle\widehat G_i(t)\big\rangle\,
\big\langle\widehat G_j(t)\big\rangle,
\label{eq:QFIM_varG}
\end{equation}
which may equivalently be written as
\begin{equation}
F_{ij}(t)
= 4\,\mathrm{Cov}\!\Big(\widehat G_i(t),\,\widehat G_j(t)\Big)
\;\equiv\;
4\Big(
\tfrac{1}{2}\big\langle\big\{\widehat G_i,\widehat G_j\big\}\big\rangle
-\big\langle\widehat G_i\big\rangle\big\langle\widehat G_j\big\rangle
\Big),
\label{eq:QFIM_cov} 
\end{equation}
with the symmetrised (anti-commutator) product ensuring $F_{ij}=F_{ji}$.

\subsection*{Keldysh representation}

For coupling deformations, the local deformation operators associated with each parameter are
\begin{equation}
\hat o_{i,H}(t)
\;\equiv\;
\int d^3x\;\partial_i\widehat{\mathcal L}_{\boldsymbol\lambda}(t,x)\Big|_H,
\qquad
\widehat G_i(t)
= -\frac{1}{\hbar}\int_0^t\!dt'\;\hat o_{i,H}(t'),
\label{eq:oi_def}
\end{equation}
generalising Eqs.~\eqref{eq:odef_S}-\eqref{eq:G_from_oH}.  Substituting into~\eqref{eq:QFIM_cov} gives the multi-parameter connected symmetrised correlator,
\begin{equation}
F_{ij}(t)
=\frac{4}{\hbar^2}
\int_0^t\!dt'\int_0^t\!dt''\;
\frac{1}{2}\Big\langle
\big\{\delta\hat o_{i,H}(t'),\;\delta\hat o_{j,H}(t'')\big\}
\Big\rangle,
\label{eq:QFIM_symm}
\end{equation}
where $\delta\hat o_i \equiv \hat o_i - \langle\hat o_i\rangle$ as before.

To express this in the Schwinger-Keldysh formalism, we introduce a CTP generating functional with \emph{independent sources for each parameter}: 
\begin{align}
Z_{\boldsymbol\lambda}[\{J_{i+},J_{i-}\}]
=&\int \mathcal{D}\phi_f\,\mathcal{D}\phi_i\,\mathcal{D}\phi_i'\;
\rho_0(\phi_i,\phi_i')
\int\displaylimits_{\phi_+(0)=\phi_i}^{\phi_+(t)=\phi_f} \!\mathcal{D}\phi_+
\int\displaylimits_{\phi_-(0)=\phi_i'}^{\phi_-(t)=\phi_f} \!\mathcal{D}\phi_-
\nonumber\\
&\times e^{\frac{i}{\hbar}\big(S_{\boldsymbol\lambda}[\phi_+]-S_{\boldsymbol\lambda}[\phi_-]\big)
+\frac{i}{\hbar}\sum_{i=1}^{M}\int_0^{t}dt'\,
\big(J_{i+}(t')\,o_i[\phi_+(t')]-J_{i-}(t')\,o_i[\phi_-(t')]\big)}.
\label{eq:CTP_Z_multi}
\end{align}
The source $J_{i\pm}(t)$ couples to the deformation operator $o_i$ on the forward~($+$) and backward~($-$) branches respectively.  By identical manipulations to Appendix~\ref{app:CTP_derivatives}, the first functional derivatives yield
\begin{equation}
\frac{\delta\ln Z_{\boldsymbol\lambda}}{\delta J_{i+}(t)}
= \frac{i}{\hbar}\,\langle o_{i+}(t)\rangle_J,
\qquad
\frac{\delta\ln Z_{\boldsymbol\lambda}}{\delta J_{i-}(t)}
= -\frac{i}{\hbar}\,\langle o_{i-}(t)\rangle_J,
\label{eq:first_deriv_multi}
\end{equation}
and the mixed second derivatives give the connected cross-branch correlators
\begin{equation}
\frac{\delta^2\ln Z_{\boldsymbol\lambda}}
{\delta J_{i-}(t')\,\delta J_{j+}(t'')}
\Bigg|_{\{J\}=0}
=\frac{1}{\hbar^2}\Big(
\big\langle\hat o_{i,H}(t')\,\hat o_{j,H}(t'')\big\rangle
-\big\langle\hat o_{i,H}(t')\big\rangle
\big\langle\hat o_{j,H}(t'')\big\rangle
\Big).
\label{eq:mixed_lnZ_multi}
\end{equation}
Setting $J_\pm=0$, the $(-)$-branch insertion is later on the Keldysh contour than the $(+)$-branch insertion, so the right-hand side is the connected Wightman function $\langle\hat o_{i,H}(t')\,\hat o_{j,H}(t'')\rangle_c$, exactly as in the single-parameter case (cf.\ Eq.~\eqref{eq:mixed_lnZ_final}).

Combining the two inequivalent mixed derivatives and using the symmetry of the double time-integral, the QFIM takes the compact Keldysh form
\begin{equation}
F_{ij}(t)
=4\,\mathrm{Re}
\int_0^{t}\!dt'\int_0^{t}\!dt''\;
\left[
\frac{\delta^2\ln Z_{\boldsymbol\lambda}}
{\delta J_{i-}(t')\,\delta J_{j+}(t'')}
\right]_{\{J\}=0}.
\label{eq:QFIM_Keldysh}
\end{equation}
This is the direct multi-parameter generalisation of Eq.~\eqref{eq:QFI_from_lnZ}.  The symmetry $F_{ij}=F_{ji}$ follows from relabelling $t'\leftrightarrow t''$ together with $i\leftrightarrow j$ under the double integral.

\subsection*{Semiclassical reduction}

In the semiclassical (Van Vleck-Gutzwiller) limit, the multi-parameter QFIM reduces to the Wigner-weighted covariance of classical action derivatives.  Following the same steps as in Section~3 and Appendix~\ref{sec:semiclassical_from_pqfi_1d}, one obtains
\begin{equation}
F_{ij}(t)
\;\approx\;
\frac{4}{\hbar^2}\,\mathrm{Cov}_{\rm sc}\!\big(\partial_i S_{\rm cl},\;\partial_j S_{\rm cl}\big)
\;=\;
\frac{4}{\hbar^2}
\Big(
\big\langle(\partial_i S_{\rm cl})(\partial_j S_{\rm cl})\big\rangle_{\rm sc}
-\big\langle\partial_i S_{\rm cl}\big\rangle_{\rm sc}\,
\big\langle\partial_j S_{\rm cl}\big\rangle_{\rm sc}
\Big),
\label{eq:QFIM_sc}
\end{equation}
where $\partial_i S_{\rm cl} = \int_0^t dt'\int d^3r\;\partial_i\mathcal L_{\boldsymbol\lambda}(\phi_{\rm cl})$ is the parametric derivative of the classical action along the trajectory $\phi_{\rm cl}$ generated by initial data $(\Phi,\Pi)$, and $\langle\cdot\rangle_{\rm sc}$ denotes the semiclassical phase-space average~\eqref{eq:sc_average_functional_cont} weighted by the initial-state Wigner functional $W_0[\Phi,\Pi]$.

Equation~\eqref{eq:QFIM_sc} is the natural multi-parameter extension of the single-parameter variance formula~\eqref{eq:F_sc_variance_field}: the diagonal elements reproduce the variance $\mathrm{Var}_{\rm sc}(\partial_i S_{\rm cl})$, while the off-diagonal elements capture the statistical correlations between different action deformations accumulated along the same classical trajectory.  This structure mirrors the classical multi-parameter Cram\'er-Rao bound, with the Wigner-weighted classical ensemble playing the role of the probability distribution over experimental outcomes.

\end{document}